\documentclass[twoside]{article}
\usepackage{fleqn,espcrc2}

\newcommand{\Sec}[1]{{\frenchspacing Section~\ref{#1}}}

\title{QCD on $\alpha$-Clusters
  \thanks{Talk presented by N. Eicker.}}

\author{N.~Eicker$^{\rm a,b}$, C.~Best\address{John von
    Neumann-Institut f\"ur Computing, c/o Research Center J\"ulich,
    D-52425 J\"ulich, Germany}, Th.~Lippert\address{Department of
    Physics, University of Wuppertal, D-42097 Wuppertal, Germany}, and
  K.~Schilling$^{\rm a,b}$}

\date{}

\begin{document}

\begin{abstract}
  It is shown that the 21264 Alpha processor can reach about 20\%
  sustained efficiency for the inversion of the Wilson-Dirac operator.
  Since fast ethernet is not sufficient to get balancing between
  computation and communication on reasonable lattice-- and
  system--sizes, an interconnection using Myrinet is
  discussed. We find a price/performance ratio
  comparable with state-of-the-art SIMD-systems for lattice QCD.
\end{abstract}

\maketitle

\section{Introduction}

The urgent need for cheap sustained compute power for lattice QCD
(LQCD) provides a strong motivation to fathom the potential of PC or
workstation clusters. It is not a long time ago that PCs and
workstations have become both speedy and cheap enough to render their
clustering in commodity networks economical, in view of local
performance, scalability and total system size. Moreover, to render
clusters efficiently one needs open source operating systems such as
Linux. The apparent success of Beowulf clusters and the tremendous
peak compute power of Alpha processors as realized in the Avalon
cluster \cite{AVALON} immediately have called the attention of the
lattice community.

We are going to investigate two different cluster approaches, both
based on Compaq Alpha processors:

One system (NICSE-TS) we have designed and benchmarked, using
state-of-the-art iterative solver codes, is a four-node cluster of 533
MHz 21164 EV56 Alpha processors, installed as a test-system at the
{\em John von Neumann-Institut f\"ur Computing} in J\"ulich/Germany
and operated under Linux.  Since QCD involves only nearest-neighbor
interaction, a mesh based connectivity appeared to be the natural
parallel architecture in order to handle the ensuing interprocessor
communication between the nodes.

Our second test-cluster (ALiCE-TS) has been designed with respect to
the experiences gained by NICSE-TS. Besides the shift to 21264 EV6
Alpha processors we are using Myrinet, a Gbit network. This promises
the interprocessor connectivity to be fast enough to compute LQCD on
Alpha clusters.  As Myrinet provides a multi-stage crossbar, we have
given up the former mesh approach. This test system again consists of
four workstations. We will show that ALiCE-TS is superior to the
``cheap'' NICSE-TS in terms of price/performance ratio by nearly a
factor of two.

The paper is organized as follows: in \Sec{TESTBED} we give the
specifications for the two variant clusters tested, \Sec{CODE_RESULTS}
describes our benchmark codes and contains some results and in
\Sec{CONCLUSION}, we give price/performance ratios.

\section{The Testbeds\label{TESTBED}}

The benchmark systems consist each of four single processor nodes
with two different generations of Alpha processors. The connectivity
is fast Ethernet and Myrinet, respectively.

\subsection{NICSE-TS}

NICSE-TS is a four-node system with fast Ethernet connectivity.  The
system is located at NIC, FZ-J\"ulich. The nodes are very similar to
the Avalon-nodes, i.e. they contain:
\begin{itemize}
\item 533 MHz 21164A Alpha microprocessors, 2 MB 3$^{\rm rd}$
  level cache, Samsung Alpha-PC 164UX motherboards
\item ECC SDRAM DIMMs (256 MB per node)
\item D-Link DFE 500 TX Ethernet cards
\item MPI based on MPIch
\end{itemize}
The main difference to Avalon is the network-setup. Where Avalon has
an all-to-all network using switches, the NICSE-TS uses a 2-D torus.
Thus we need four Ethernet cards per node where Avalon only employs
one. On the other hand we do not need any switch.  We expect, that the
network performance scales to a large number of nodes for nearest
neighbor communication.  All-to-all communication can be achieved by
the routing capabilities of the Linux kernel.

\subsection{ALiCE-TS}

ALiCE-TS is a four-node cluster with switched Myrinet connectivity.
This system is hosted at Wuppertal University. It includes:
\begin{itemize}
\item 466 MHz 21264 Alpha microprocessors,\\ 2 MB 2$^{\rm nd}$ level
  cache, Compaq DS10 motherboards
\item ECC SDRAM DIMMs (128 MB per node)
\item 64-bit 33MHz Myrinet-SAN/PCI interface
\item MPI based on Myrinet GM library
\end{itemize}
ALiCE-TS has been purchased as prototype system for the design of the
128 node Wuppertal Alpha-Linux-Cluster Engine (ALiCE).

\section{QCD Benchmarks and Results\label{CODE_RESULTS}}

The computational key problem of LQCD is the---very often
repeated---inversion of the Dirac matrix. It has been shown in
\cite{FROMMER}, that such systems are most efficiently solved by
Krylov subspace methods like BiCGStab.  State-of-the-art is the
application of parallel local lexicographic preconditioning within
BiCGStab \cite{FISCHER}.

The results of this paper's benchmarks are based on two codes:

{\bf BiCGStab} is a sparse matrix Krylov solver with regular memory
access, where computation and communication proceed in an alternating
fashion.  In this case, DMA capabilities of the communication cards
are not exploited.

{\bf SSOR} is the same solver but with local-lexicographic SSOR
preconditioning. The SSOR process leads to rather irregular memory
access and extensive integer computations. This code is very sensitive
to the memory-to-cache bandwidth. Since communication overlaps with
computation, DMA can be exploited.

Both codes are written in {\sf C} and compiled under the GNU {\sf
  egcs-1.1.2 C} compiler. Timing was done with {\tt MPI\_Wtime}. For
both codes there exist two versions:
\begin{enumerate}
\item To test single node performance, the code runs without
  communication operations, otherwise carrying out exactly the same
  operations as the following parallel version.
\item On the 4-node test machines, the physical system is laid out in
  a 2-D fashion, consequently, communication is carried out along two
  dimensions, namely $z$- and $t$-directions. Assuming
  $N_{proc}=N_z\times N_t$ processors, the global lattice is divided
  in $N_t$ slides in $t$-direction where every slide consists of
  $N_z$ slides in $z$-direction.
\end{enumerate}
In the sequel, we are going to employ a local lattice of size
$16^2\times 4\times 8$ on $2\times 2$ processors such that we emulate
a realistic $16^3\times 32$ system on $4\times 4$ processors.

\subsection{Single-node results}

The basic operation in the iterative solution of the Dirac matrix is
the product of a SU(3) matrix with two color vectors.  The average
number of flops per matrix vector operation is $N_{flop}=171$.  The
number of complex words to get from memory in order to
carry out this process is $N_{cwords}=(9+2\times 12)$ leading to
$N_{bytes}=528$ bytes for double precision arithmetics.  Therefore we
expect the maximal performance that can be reached for a single node
to be limited by
\begin{displaymath}
P_{max}=\frac{B}{N_{byte}}N_{flop}=\left\{
\begin{array}{ll}
97 &\mbox{MFlops (UX)}\\
420&\mbox{MFlops (DS10)}\\
\end{array}\right.
\end{displaymath}
in a steady state of computation and data flow, given a maximal memory
bandwidth of 300 and 1300 MB/sec, respectively.  Note that our
problem size is chosen to be larger than the available caches. The
real performances will be smaller due to BLAS-1 and BLAS-2 operations
within BiCGStab.

\begin{table}[htb]
\begin{center}
\begin{tabular}{|l||l|l||l|l|l|l|}
\hline
&\multicolumn{2}{c||}{double prec.}&\multicolumn{2}{c|}{single prec.}\\
\hline
\hline
Benchmark & UX & DS10 & UX & DS10 \\
\hline
\hline
BiCGStab & 82 & 166 &   116 &  232 \\
SSOR     & 57 & 115 &   90  &  182  \\
\hline
\end{tabular}
\end{center}
\caption{One processor benchmark. Numbers in MFlops.\label{SINGLENODE}}
\end{table}

Table~\ref{SINGLENODE} shows that, on the UX board, the performance of
BiCGStab comes close to the limiting value given, while the DS10
performance deviates by more than a factor of 2 from the 
estimate. The local lattice size presumably is too small to lead to
saturation of the bandwidth for the DS10\footnote{The STREAMS
  benchmark \cite{STREAMS} gives a real bandwidth of 580 MB/sec
  instead of the theoretical value of 1300. This difference explains
  the factor of two.}. However, as a main result, we find that the
improvement in performance going from the 533 MHz Alpha 21164 to the
466 MHz Alpha 21264 chip is around a factor of two, using identical
codes.  Furthermore, the SSOR preconditioner with irregular memory
access is, as has been expected, less effective than the simple
BiCGStab.

\subsection{Four-node results}

The impact of interprocessor communication for both connectivities
is determined on the four-node testbed systems.

\begin{table}[htb]
\begin{center}
\begin{tabular}{|l||l|l||l|l|l|l|}
\hline
&\multicolumn{2}{c||}{double prec.}&\multicolumn{2}{c|}{single prec.}\\
\hline
\hline
Benchmark & UX & DS10 & UX & DS10 \\
\hline
\hline
BiCGStab & 32 & 130 &   54 &  201 \\
SSOR     & 30 & 100 &   53 &  164  \\
\hline
\end{tabular}
\end{center}
\caption{Four processor benchmark. Numbers in MFlops.\label{FOURNODE}}
\end{table}

As shown in Table~\ref{FOURNODE}, the results for the fast Ethernet mesh
(denoted by UX) are disappointing. The performance of both codes, SSOR
and BiCGStab, is reduced by more than a factor of two compared to the
single node result. The main degradations are due to the massive
protocol overhead forcing the processor into administration instead
of computation.  User-level networking interfaces promise to
circumvent this problem in the near future, but are currently not
available for our configuration.

It is satisfying to see, by comparing Tables~\ref{SINGLENODE} and
\ref{FOURNODE}, that the Alpha 21264-Myrinet system (denoted by DS10)
with Myrinet GM library has a communication loss in the range of only
10 to 20~\%. We expect a further considerable improvement of these
results by employing software with reduced protocol stack like SCore
\cite{SCORE} or ParaStation \cite{PARA}.

\section{Conclusion\label{CONCLUSION}}

Comparing price/performance ratios we arrive at the following
estimates: An Alpha 21164 system, connected in a fast Ethernet mesh,
would---as an optimistic estimate---lead to a 4 GFlops device
(sustained) for 128 processors with a price of about 80 k\$ per
GFlops.

A 128 processor DS10 Alpha-Linux-Cluster connected by Myrinet,
however, promises to reduce costs to 40 -- 50 k\$ per GFlops
(estimated from list prices as of July 1999) and is therefore in the
range of state-of-the-art dedicated QCD machines.


\begin{thebibliography}{99}
\vspace{-0.125cm}
\frenchspacing
%
\bibitem{AVALON} http://cnls.lanl.gov/avalon.
%
\bibitem{FROMMER} A.~Frommer, V.~Hannemann, B.~N\"ockel, Th.~Lippert, and
  K.~Schilling: Int. J. of Mod. Phys. C Vol.~5 No.~6 (1994) 1073.
%
\bibitem{FISCHER} S.~Fischer, A.~Frommer, U.~Gl\"{a}ssner, Th.~Lippert,
  G.~Ritzenh\"{o}fer, and K.~Schilling: Comp. Phys. Comm.~98 (1996)
  20.
%
\bibitem{STREAMS} http://www.cs.virginia.edu/stream
%
\bibitem{SCORE} http://pdswww.rwcp.or.jp/dist/score.
%
\bibitem{PARA} http://ParaStation.ira.uka.de.
%
\nonfrenchspacing
\end{thebibliography}
\end{document}